\documentclass{article}
\usepackage[utf8]{inputenc}
\DeclareUnicodeCharacter{2009}{\,}
\usepackage{float}
\usepackage{graphicx}
\usepackage{caption}
\usepackage{multirow}
\usepackage{comment}
\usepackage{tabularx}
\usepackage{subcaption}
\usepackage{amsmath}
\usepackage{authblk}

\title{De Novo Design of SIK3 Inhibitors via Feedback-Driven Fine-Tuning of Seq2Seq-VAE}

\author[1,2]{ShahZeb Khan}
\author[1]{Chiara Pallara}
\author[2]{Barbara Monti}
\author[1,*]{Alexis Molina}

\affil[1]{Nostrum Biodiscovery S.L., Av. de Josep Tarradellas, 8-10, 3-2, 08029, Barcelona, Spain}
\affil[2]{Department of Pharmacy and Biotechnology, Alma Mater Studiorum–University of Bologna, Bologna, Italy}
\begin{document}

\maketitle

\begin{abstract}
Alzheimer’s disease (AD), a progressive neuro-degenerative disorder, currently lacks effective therapeutic strategies that can modify disease progression. Recent studies have highlighted the circadian rhythm’s critical role in AD pathophysiology, implicating circadian clock kinases, such as the Salt-Inducible Kinase 3 (SIK3), as promising therapeutic target. Generative AI models have surpassed traditional methods of drug discovery, untapping the vast unexplored chemical space of drug-like molecules. We present a sequence-to-sequence Variational Autoencoder (Seq2Seq-VAE) model guided by an Active Learning (AL) approach to optimize molecular generation. Our pipeline iteratively guided a pre-trained Seq2Seq-VAE model towards the pharmacological landscape relevant to SIK3 using a two-step framework, an inner loop that iteratively improves physio-chemical properties profile, drug likeliness and synthesizability, followed by an outer loop that steer the latent space towards high-affinity ligands for SIK3. Our approach introduces feedback-driven optimization without requiring large labeled datasets, making it particularly suited for early-stage drug discovery in underexplored therapeutic targets. Our results demonstrate  the model's convergence toward SIK3-specific small molecules with desired properties and high binding affinity. This work highlights the use of generative AI combined with AL for rational drug discovery that can be extended to other protein targets with minimal modifications, offering a scalable solution to the molecular design bottleneck in drug design.
\end{abstract}

\section*{Introduction}
Dementia poses a formidable challenge to global healthcare, affecting approximately 55.2 million individuals worldwide, as of 2022, a number projected to double every two decades until 2050 \cite{Zhang2024-km}. Alzheimer's disease (AD), a neuro-degenerative disorder, the primary cause of dementia, represents a growing global burden with few effective treatments. Recent studies emphasize that AD’s intricate etiology, involving aging, genetics, and environmental factors, exacerbates its global impact, necessitating urgent therapeutic advancements \cite{Zhang2024-km}. As of early 2023, 187 clinical trials were underway, with 78\% focused on disease-modifying therapies instead of targeting specific proteins \cite{Cummings2023-dw}. Current therapeutic developments are failing, with over 99.6\% of small molecule drugs failing in clinical trials. Even with recent approvals of amyloid-targeting antibodies, the AD drug development pipeline continues to face significant hurdles, underscoring the need for novel approaches to achieve meaningful therapeutic outcomes \cite{Fang2020-ml}. AD is characterized by the accumulation of amyloid-beta (A$\beta$) plaques and the hyper-phosphorylation of tau protein, leading to Neurofibrillary Tangles (NFTs) \cite{Zhang2024-km}. Despite ongoing efforts, the complex molecular mechanisms of AD have restrained the development of effective therapeutics. Several studies have reported a close link between the disruption of circadian rhythms and AD \cite{Homolak2018-tu, Maiese2020-rl}. This has increased interest in exploring circadian rhythm-based innovative options for AD treatment \cite{Hoyt2022-du}. Reports from animal studies suggest that AD models, particularly genetically modified mice that over-expression of Amyloid Precursor Protein (APP) or A$\beta$, exhibit significant disruptions in their circadian rhythms, affecting sleep patterns, movement, and body temperature regulation \cite{Wisor2005-mx}. Similar observations have been made in humans, where more pronounced circadian disturbances include increased fragmentation, shifts in phase, and reduced amplitude \cite{Videnovic2014-yu, Ahmad2023-kt}. Salt inducible kinase 3 (SIK3), a member of the SIKs subfamily of kinases, has been widely reported to be highly expressed in the brain and linked to destabilization of the Period Circadian regulator 2 (PER2), a major clock gene. SIK3 plays a key role in disruption of circadian rhythms, as it facilitates the destabilization of PER2, either directly or indirectly, through phosphorylation-dependent mechanisms \cite{Hayasaka2017-do}. Given its role in disrupting circadian rhythms, SIK3 represents a promising target for small molecules aimed at restoring normal circadian function. As SIKs are involved in a multitude of physiological functions, there has been a great interest in SIK’s specific inhibitor development for pharmacological interventions, for instance, SIK2 or SIK2/3 specific inhibitors have been designed to treat osteoporosis and ovarian cancer. In addition, Galapagos, a Belgium-based pharmaceutical company, has developed SIK inhibitor (GLPG3312) with considerable specificity profile, as a treatment option for autoimmune and inflammatory diseases \cite{Oster2024-aa, Temal-Laib2024-gf}. Even though significant progress has been made in this direction, SIK3 has never been used as a drug target for AD, moreover, no SIK inhibitors reported to date can pass the blood-brain barrier (BBB). The growing focus on non-amyloid, non-tau targets, such as those related to circadian rhythm, underscores the potential of underexplored targets like SIK3 in AD therapeutic development. \par 

The drug discovery process requires a comprehensive approach to achieve desired therapeutic outcomes. This involves the creation and selection of potential ligands, followed by rigorous testing to evaluate their interactions with specific targets. Recent advancements in Artificial Intelligence (AI) has demonstrated its potential to significantly enhance various tasks involved in the drug discovery process, showcasing better performance than traditional methods \cite{Schneider2020-hr}. These AI models enable ligand optimization, predicting ADME properties, and aiding in the discovery of novel drug targets. A more promising application of AI lies in the domain of \textit{de novo} drug design, which focuses on leveraging generative AI models to generate entirely new chemical compounds (NCCs) with desired properties from scratch. The number of potential drug-like molecules that are synthetically accessible and have a molecular weight less than 500 daltons is vast, estimated at approximately $10^{60}$ distinct entities. Nevertheless, only a fraction, approximately $10^8$, has been discovered and synthesized, indicating that over 99\% of the chemical space remains unexplored \cite{Virshup2013-vv}. Experimentally exploring this immense chemical space is both costly and inefficient, highlighting the key role of generative models in this direction. Both industrial and academic researchers have made significant strides in drug development utilizing AI tools, with some focusing on AD/ADRD (AD-related dementia) \cite{Cheng2022-dw}. For example, a generative model was trained to generate BACE1 inhibitors using extensive existing BACE1 binding affinity data \cite{Xie2024-jf}. Similarly, Zhavoronkov \textit{et al}., 2019, validated the practical utility of generative models through the design of potent DDR1 kinase inhibitors \cite{Zhavoronkov2019-vk}. \par

To date, 158 AI-driven drug candidates have entered discovery or preclinical stages across multiple diseases \cite{noauthor_2023-av}. AD remains especially challenging because of its complex pathophysiology and the limits of traditional discovery pipelines \cite{Huang2023-aq}. Here we present a data-efficient, structure-aware \textit{de novo} design approach that couples a Seq2Seq-VAE with a two-loop active-learning curriculum to focus generation on SIK3, a target implicated in AD-related circadian disruption, while maintaining central-nervous-system feasibility. The inner loop enforces conservative physio-chemical and developability constraints, including drug-likeness, synthetic accessibility, and BBB-relevant criteria, to ensure therapeutically plausible profiles. The outer loop applies structure-guided selection by docking to SIK3 and prioritizing molecules that satisfy predefined interaction hypotheses at the kinase hinge, thereby steering the model toward SIK3-compatible chemotypes. Despite scarce target-specific starting data, the curriculum progressively enriches the latent space for candidates that meet both property and docking filters while retaining scaffold diversity. All evidence in this study is \textit{in silico} and reflects docking-based enrichment and pose stability rather than measured potency, but the results demonstrate that a lightweight, reproducible procedure can translate limited prior knowledge into focused candidate sets. The same framework is readily extensible to other underexplored targets, positioning it as a scalable strategy for \textit{de novo} molecular design in data-limited settings. \par

\begin{figure}[ht]
\centering
\includegraphics[width=\linewidth]{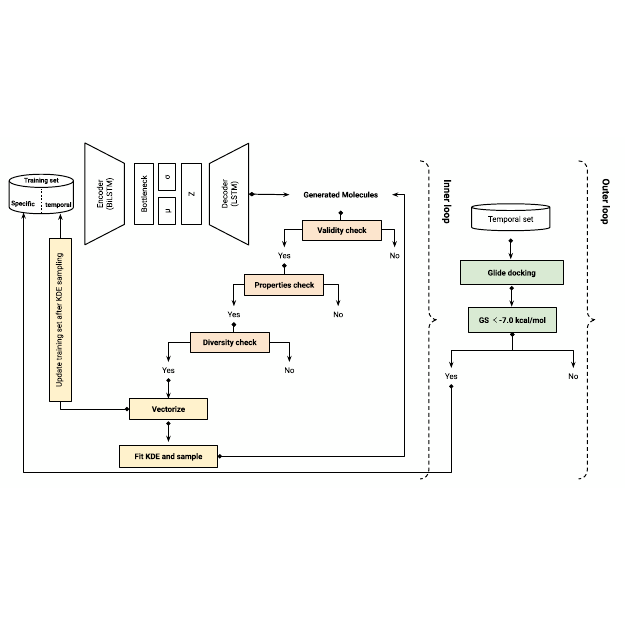}
\captionsetup{justification=justified, singlelinecheck=false}
\caption{Generation of SIK3-specific molecules by AL workflow. The workflow involves two nested iterative processes: an inner loop focused on optimizing physio-chemical properties and an outer loop focused on optimizing the binding affinity for SIK3. During each inner loop, new molecules are generated from KDE-sampled latent space and filtered based on physio-chemical properties (Table 1). The resulting filtered molecules are used to enrich the temporal-specific set during the inner loops. Upon completion of a specified number of inner loops, an outer loop filters the molecules from the temporal-specific set based on their docking score ($\leq -7.0 \, \text{kcal/mol}$ for cycles 1--4 and $\leq -7.5 \, \text{kcal/mol}$ for cycles 5--8). The filtered molecules are then transferred from the temporal-specific set to the permanent-specific set. After a specified number of outer loops, all generated molecules in the permanent-specific set undergo further filtration for SIK kinase specificity.}
\label{fig:workflow}
\end{figure}

\begin{table}[ht]
\centering
\captionsetup{justification=justified, singlelinecheck=false}
\caption{Cutoff criteria for inner and outer loops in the AL workflow. The Quantitative Estimate of Drug-likeliness (QED) cutoff was progressively tightened from $\geq 0.5$ to $\geq 0.6$ across inner loop cycles to balance exploration and optimization.}
\label{tab:cutoffs}
\renewcommand{\arraystretch}{1.3} % Increase row spacing by 30%
\begin{tabularx}{\textwidth}{|p{2.5cm}|X|p{3.5cm}|}
\hline
\textbf{Loop} & \textbf{Property} & \textbf{Cutoff} \\
\hline
\multirow{8}{*}{Inner loop} & Validity & + \\
& Quantitative Estimate of Drug-likeliness & $\geq 0.5$ to $\geq 0.6$ \\
& Synthetic Accessibility & 1 to 6 \\
& Molecular Weight & $\leq 500$ \\
& logP & 0 to 4 \\
& TPSA & $\leq 90$ \\
& Number of Oxygen and Nitrogen atoms & $\leq 4$ \\
& Hydrogen Bond Donor & $\leq 3$ \\
\hline
\multirow{4}{=}{Outer loop} & \multirow{4}{=}{Glide Docking score} & $\leq -7.0 \, \text{kcal/mol}$ (cycles 1--4) \newline $\leq -7.5 \, \text{kcal/mol}$ (cycles 5--8) \\
\hline
\end{tabularx}
\end{table}

\section*{Results}
A Seq2Seq-VAE was trained on ~670,000 molecules from ChEMBL to learn general molecular syntax. This pretrained model was fine-tuned through a two-tiered active learning (AL) framework to bias generation toward molecules with desired pharmacological profiles for SIK3.  The inner loop, aimed at improving the physio-chemical properties profile, and an outer loop, a high level optimization of molecules by prioritization based on lower molecular docking score, guiding the latent space to a chemical space of high-affinity molecules for SIK3. The AL workflow is summarized in Figure 1. \par

\subsection*{SIK3-Specific Molecule Generation via AL Workflow}
\subsubsection*{AL Inner Loop}
At the end of each inner loop cycle of the AL workflow, molecules were generated and rigorously evaluated for validity, Quantitative Estimate of Drug Likeliness (QED), and Synthetic Accessibility (SA). An additional filtering step was integrated to retain only molecules that can permeate the BBB using physio-chemical descriptors reported by Gupta \textit{et al}., 2019 \cite{Gupta2019-zu} including molecular weight (MW), logarithm of the partition coefficient (logP, a measure of lipophilicity), number of hydrogen bond donors (HBD), topological polar surface area (TPSA, reflecting molecular polarity), and the count of oxygen and nitrogen atoms, specific cutoffs detailed in Table 1. The KDE-based sampling strategy, at bandwidth 0.3, efficiently guided the model towards generating desired molecules. The AL workflow comprised eight inner loops, employing progressively stricter cutoffs (initially $ \text{QED} \geq 0.5 $, Tanimoto similarity 0.8; later $ \text{QED} \geq 0.6 $, Tanimoto similarity 0.7) to adapt to increasing complexity. Notably, the number of valid molecules decreased from 91\% (6832/7500) to 82\% (6209/7500) in the first four cycles, while molecules meeting physio-chemical properties criteria rose from 56\% (3849/6832) to 80.7\% (5015/6209). \par

With stricter cutoffs, valid molecules dropped from 76.3\% (11448/15000) to 71.6\% (10746/15000), yet molecules passing the physio-chemical filters increased from 79\% (9054/11448) to 85.4\% (9187/10746). The inner loop cycles within the AL workflow demonstrates its efficacy in enhancing the generation of high-quality, BBB-permeable molecules with desired physio-chemical properties. The temporal-specific set was updated per cycle. \par

\subsubsection*{AL Outer Loop and Docking-Based Selection}
An outer loop, at the end of every 5 inner loop cycles, was executed as a higher-level optimization and selection process. Molecules accumulated in the temporal-specific set during inner loops were filtered based on glide docking scores with SIK3 and the presence of a critical Ala145 hydrogen bond, essential for SIK3 inhibitor stability. Only molecules exhibiting a docking score of $\leq -7.0 \, \text{kcal/mol}$ and an Ala145 hydrogen bond were prioritized and transferred to the permanent-specific set. Over the first four outer loops, this dual-criterion filtering expanded the permanent-specific set from 103 to 1,218 molecules, a 12-fold increase. The AL workflow significantly improved the model’s ability to generate SIK3-binding molecules, increasing the proportion of molecules meeting both criteria from 6.0\% to 16.7\%, as shown in Figure 2a. \par

In the subsequent four outer loops, the glide docking score cutoff was tightened to $\leq -7.5 \, \text{kcal/mol}$ to further prioritize ligands. This adjustment led to the permanent-specific set growing from 1,218 molecules to 3,445 molecules, with the percentage of molecules meeting the strict criteria increasing from 6.3\% to 8\%, demonstrating continued improvement in the AL workflow’s ability to generate potent SIK3-binding molecules coupled with desired physio-chemical profiles, presented in Figure 2b. \par

\begin{figure}[ht]
\centering
\begin{subfigure}[t]{0.48\textwidth}
    \centering
    \includegraphics[width=\linewidth]{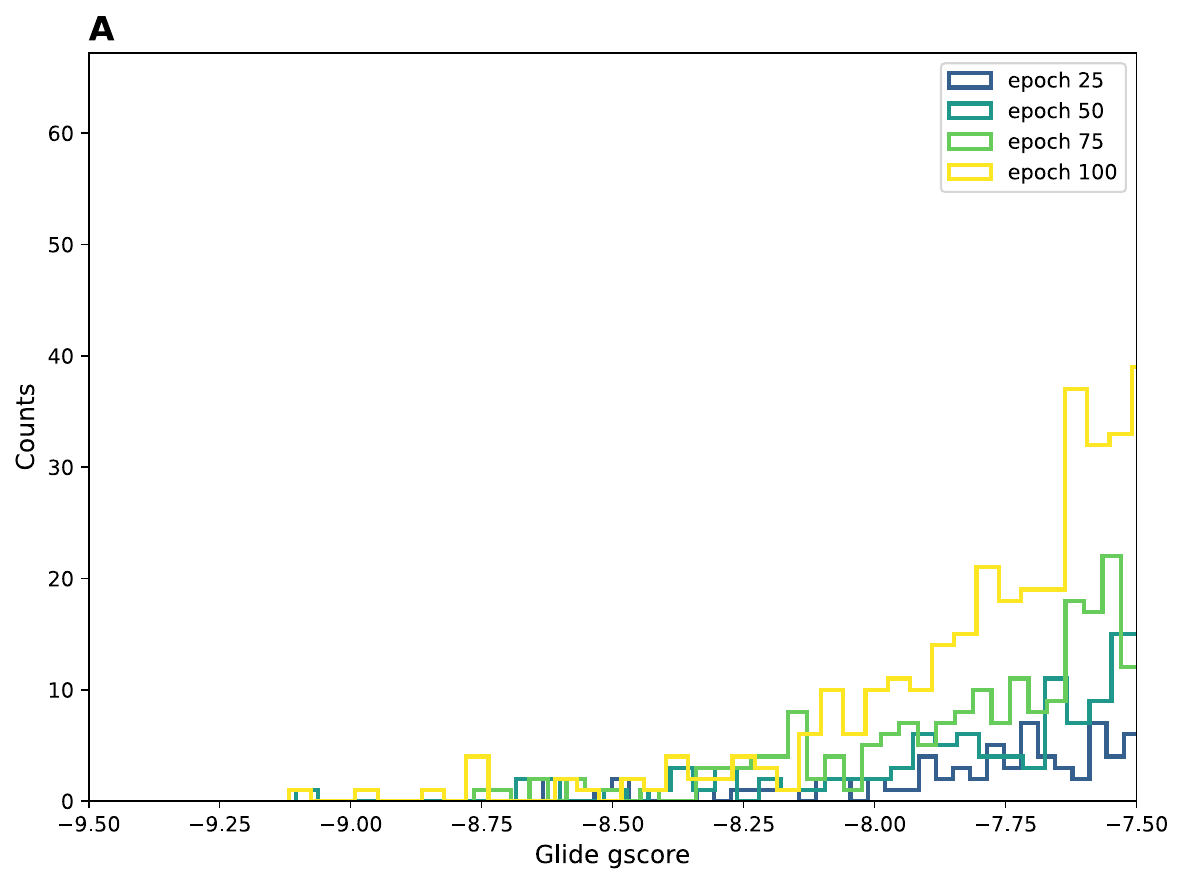}
    \captionsetup{justification=justified, singlelinecheck=false}
    \label{fig:step_histogram_25_50_75_100}
\end{subfigure}\hfill
\begin{subfigure}[t]{0.48\textwidth}
    \centering
    \includegraphics[width=\linewidth]{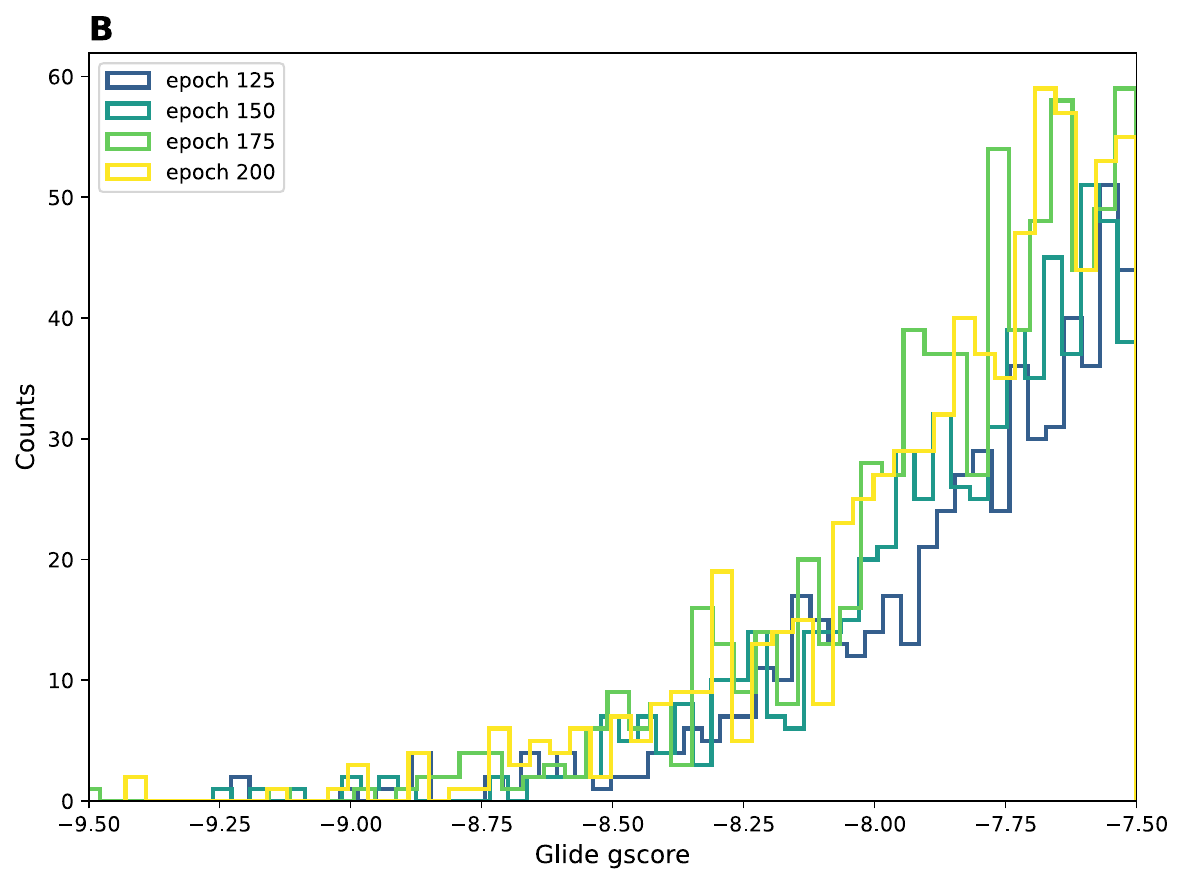}
    \captionsetup{justification=justified, singlelinecheck=false}
    \label{fig:step_histogram_125_150_175_200}
\end{subfigure}
\caption{Distribution of Glide docking scores for molecules selected during outer active learning steps at thresholds of $-7.0 \, \text{kcal/mol}$ (a) and $-7.5 \, \text{kcal/mol}$ (b).}
\label{fig:gscore_combined}
\end{figure}

\subsubsection*{AL Scaffold Diversity} 
The active learning (AL) workflow effectively enriched the temporal-specific dataset during inner loop cycles and the permanent-specific set during outer loop cycles, demonstrating the model’s robust learning capacity. This trend reflects improved alignment between the latent space distribution and desired molecular properties across inner loop iterations. To ensure the generation of diverse molecular scaffolds and prevent mode collapse, where the model repeatedly generates similar molecules, neglecting other viable chemical space regions, scaffold diversity was monitored using the Murcko Scaffold framework. The Murcko Scaffold gets the core ring structures and linkers of a molecule by removing side chains, thus focusing on its essential scaffold. Scaffold diversity is calculated as the fraction of unique Murcko scaffolds relative to the total number of generated molecules, providing a quantitative measure of structural variety. Diversity remained robust in the first four inner loop cycles, decreasing slightly from 0.81 in the first cycle to 0.73 in the fourth, confirming the model’s ability to produce diverse molecules with desired properties. A similar trend in scaffold diversity was observed in the first four outer loop cycles, with a decline from 0.91 at epoch 25 to 0.73 at epoch 100, indicating a shift toward structurally optimized scaffolds associated with high docking scores. 

In the subsequent four inner loop cycles, scaffold diversity followed a comparable trend, decreasing from 0.69 in the fifth cycle to 0.65 in the eighth, while maintaining robust structural diversity. Similarly, in the later four outer loop cycles, scaffold diversity decreased from 0.70 at epoch 125 to 0.61 at epoch 200, reflecting continued optimization of high-affinity scaffolds. Figure 3 provides detailed insights into the evolution of scaffold diversity across inner and outer loop cycles, illustrating the balance between diversity and property optimization. Figure 3 provides detailed insights into the evolution of scaffold diversity of molecule across inner and outer loop cycles. \par
\[
D_{\text{scaffold}} = \frac{N_{\text{scaffolds}}}{N_{\text{compounds}}}
\]
Where \textit{N}scaffolds is the number of unique scaffolds, and \textit{N}compounds is the total number of compounds. 

\begin{figure}[ht]
\centering
\includegraphics[width=\linewidth]{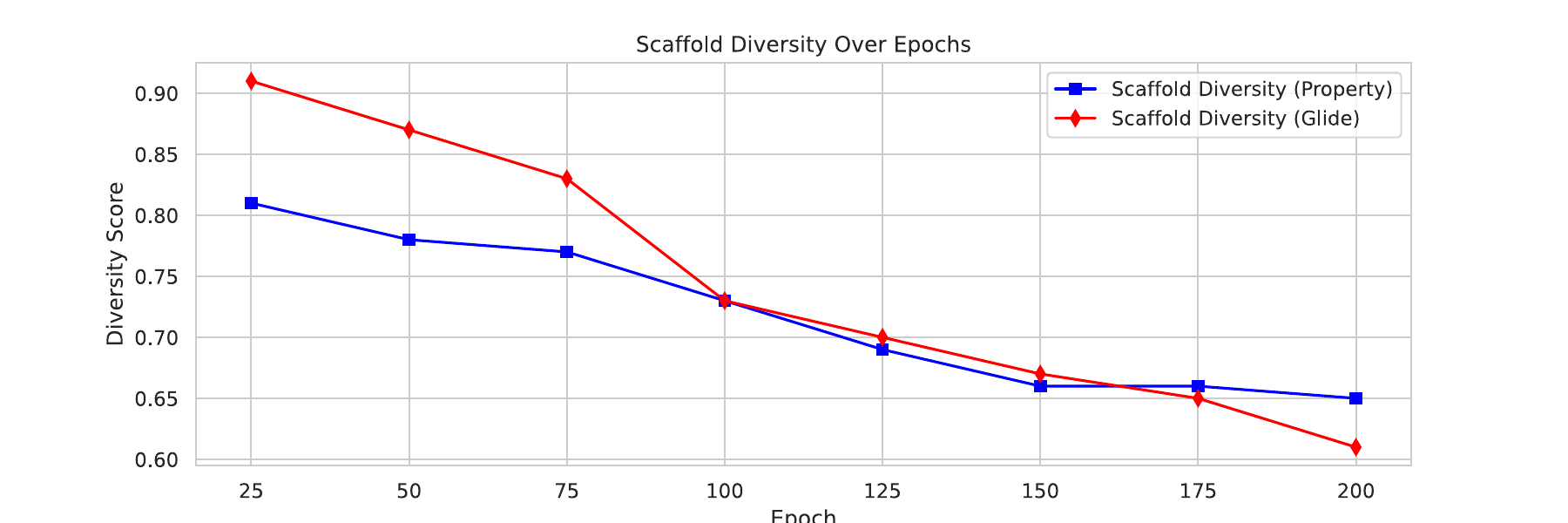}
\captionsetup{justification=justified, singlelinecheck=false}
\caption{Murcko based scaffold diversity over epochs for property-filtered and Glide-filtered SMILES. The plot shows the scaffold diversity scores for property-filtered (blue, solid line) and Glide-filtered (red, solid line) SMILES generated by the Seq2Seq-VAE model, with a general decline in diversity as training progresses, indicating a shift toward more structurally relevant latent space}
\label{fig:scaffold_plot}
\end{figure}

\subsubsection*{AL Overall Workflow Efficacy}
Overall, the finetuning process validate the effectiveness of the KDE-based AL workflow in refining the latent space representation, enhancing molecular property alignment, and improving target-specific binding potential while maintaining scaffold diversity. The increasing number of molecules passing both physio-chemical and docking-based filters highlights the ability of the iterative AL framework to drive the Seq2Seq-VAE model toward more functionally relevant chemical space. Figure 4 provides a UMAP visualization of the molecules generated over the epochs. The UMAP visualization shows the generated molecules efficiently exploring relevant chemical space, the molecules generated over the inner and outer loops throughout the AL workflow is provided in Table 2. \par

\begin{figure}[ht]
\centering
\includegraphics[width=\linewidth]{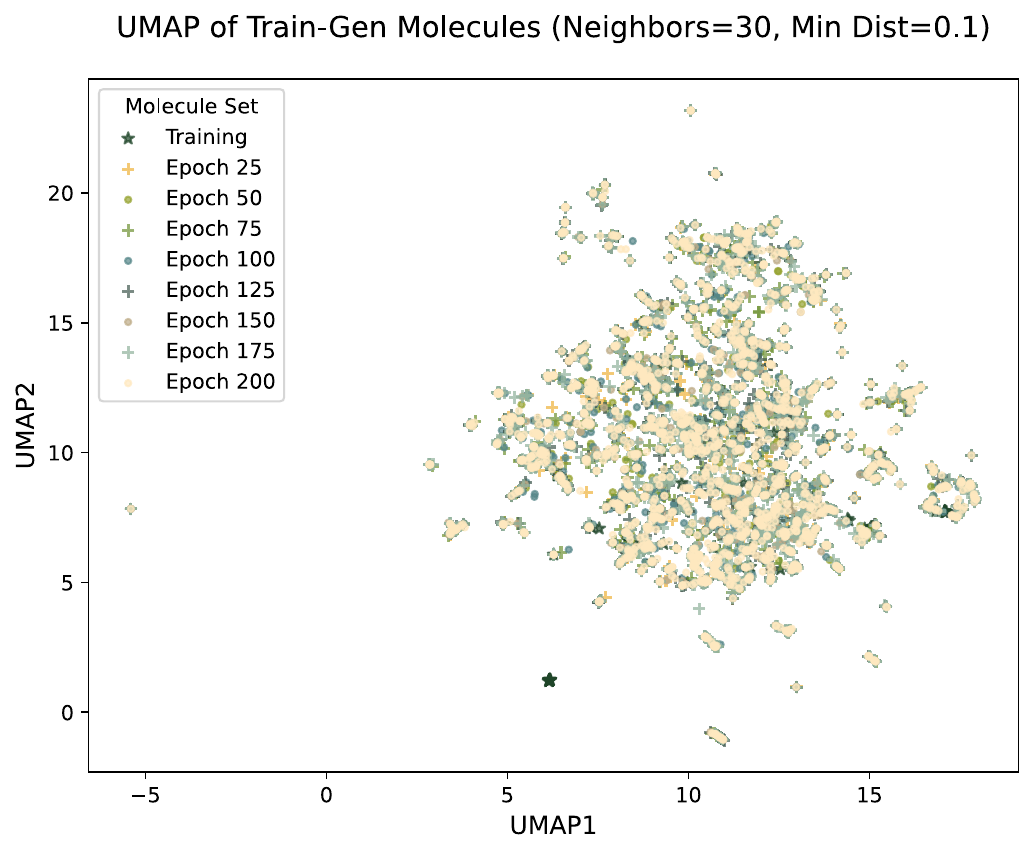}
\captionsetup{justification=justified, singlelinecheck=false}
\caption{UMAP visualization of the chemical space explored by the Seq2Seq-VAE model. Molecules from the initial training set are colored green, while candidate molecules exhibiting desired physio-chemical properties and high binding affinity for SIK3 are colored according to the legend. The plot demonstrates successful fine-tuning of the model for SIK3-specific molecule generation.}
\label{fig:UMAP}
\end{figure}

\subsubsection*{SIK family specificity and final selection}
The SIK kinases harbor a Threonine at position 142 rendering the back pocket large and accessible while other kinases, for example, the AMPK family members have large residues at this position (Methionine or Leucine) allowing for Thr142 targeting for SIK family selectivity. As reported by Galapagos in their lead optimization of GLPG3312 for SIK3 selectivity, replacement of methoxy groups difluoromethoxy moiety resulted in a loss of activity for AMPK. This loss of activity is attributed due to the presence of Methionine as a gatekeeper residue unlike Threonine in SIK kinases. Moreover, a hydrogen bond between the difluoromethoxy moiety and the threonine gatekeeper in the SIK family allows for SIK specificity of molecules \cite{Oster2024-aa, Temal-Laib2024-gf}. \par

The molecules generated and filtered at each outer loop were further filtered for the presence of Thr142 hydrogen bond to retain molecules with potential selectivity for SIK kinases. Molecules exhibiting the presence of Ala145 hydrogen bond critical for stability and Thr142 hydrogen bond critical for SIK family selectivity were further analyzed in a dynamic cell like environment using the Molecular Dynamics Simulations. \par

\subsection*{All Atoms Molecular Dynamics Simulations}
The Root Mean Square Fluctuations (RMSF) of protein residues and Root Mean Square Deviation (RMSD) over time was calculated for the protein backbone (blue line), ligand (orange line), and ligand binding site (green line) residues to assess structural stability during the molecular dynamics simulation. In addition to RMSD analysis, two distinct interactions were monitored: LIG-A85 (orange line), key interaction for ligand stability and LIG-T82 (green line), an interaction crucial for SIK family specificity. The Ala145 and Thr142 were renumbered to 85 and 82 and will be referred to as such afterwards. The protein-ligand interaction maps are provided in Figure 5.
\begin{figure}
    \centering
    \includegraphics[width=1\linewidth]{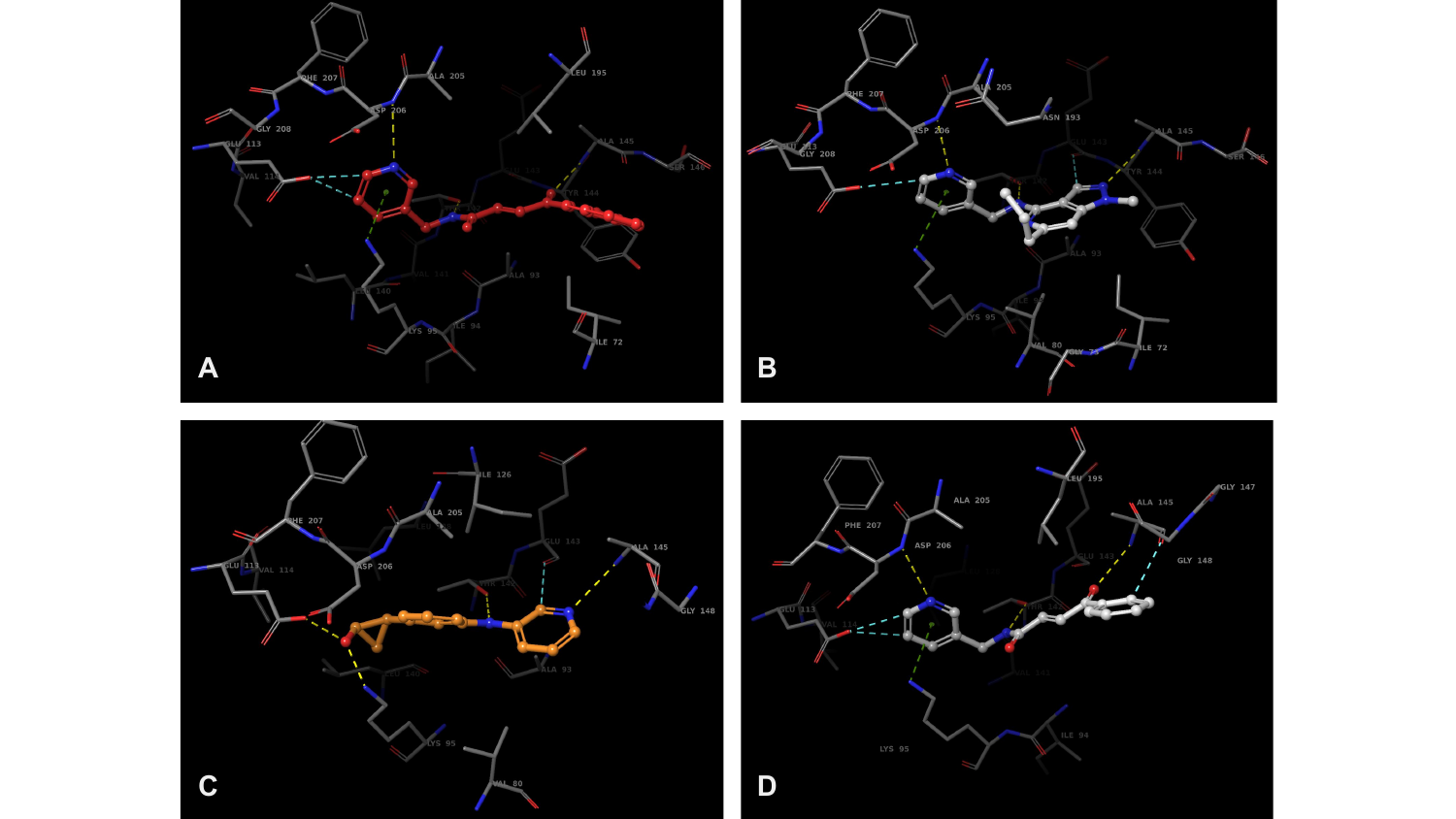}
    \caption{Protein-Ligands interactions maps highlighting the key interactions between ligand and binding site residues. (A) SIK3-1030, (B) SIK3-1459, (C) SIK3-2913 and (D) SIK3-3481}
    \label{fig:enter-label}
\end{figure}

\subsubsection*{Seq2Seq1030}
The RMSD of the protein backbone showed an initial spike, after approximately 50 ns, the RMSD stabilized around 1.5 - 2.0 Å, with periodic fluctuations reflecting moderate conformational changes. Similarly, the Seq2Seq1030 RMSD stabilized at a lower value (~1.0 - 1.6 Å), suggesting that the ligand's conformation is more constrained by its interaction with the protein. Notably, the RMSD of the ligand binding site residues remained consistently low, stabilizing near 0.7 - 1.2 Å, indicating high rigidity and minimal deviation from the reference structure. This suggests that the binding site residues maintain a stable conformation, which is crucial for maintaining the protein-ligand interaction. Overall, the system reached a dynamic equilibrium after ~50 ns, with the protein, ligand, and ligand binding site residues exhibiting characteristic patterns of stability and flexibility. The RMSD plot is presented in Figure 6.
The hydrogen bond distance comparison plot (Figure 7) illustrates the temporal fluctuations in the distances between key residues involved in hydrogen bonding interactions over the course of the simulation. Both interactions exhibit dynamic behavior, with distances fluctuating within a range of approximately 2.8 - 3.4 Å throughout the simulation. Notably, the LIG-A85 interaction shows more pronounced peaks, indicating occasional disruptions in the hydrogen bond, whereas the LIG-T82 interaction appears slightly more stable, with fewer excursions to larger distances. These fluctuations suggest that both interactions are dynamic but maintain overall stability, consistent with typical hydrogen bonding patterns observed in bio-molecular systems.

\begin{figure}
    \centering
    \includegraphics[width=1\linewidth]{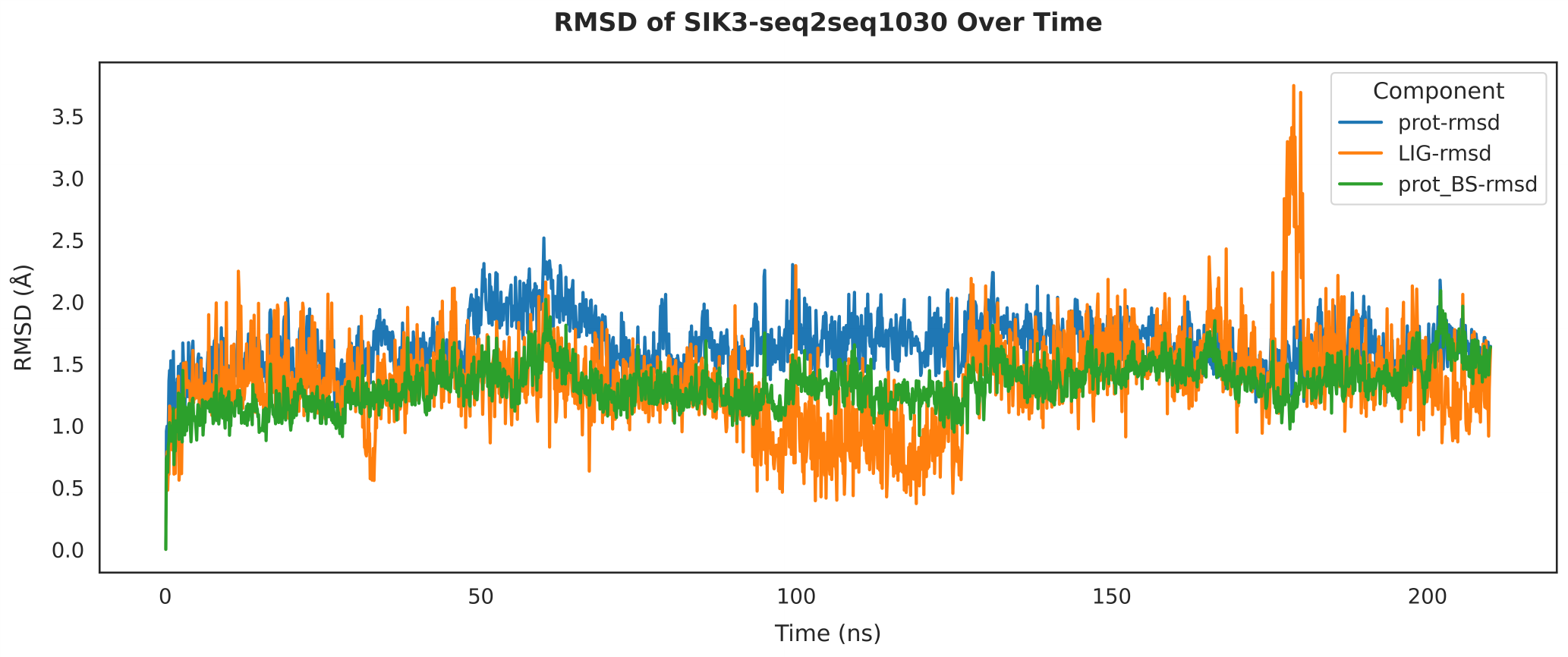}
    \caption{The RMSD plot of protein backbone (blue), Seq2Seq1030 ligand (orange), and ligand binding site (green).}
    \label{fig:RMSD}
\end{figure}
\begin{figure}
    \centering
    \includegraphics[width=1\linewidth]{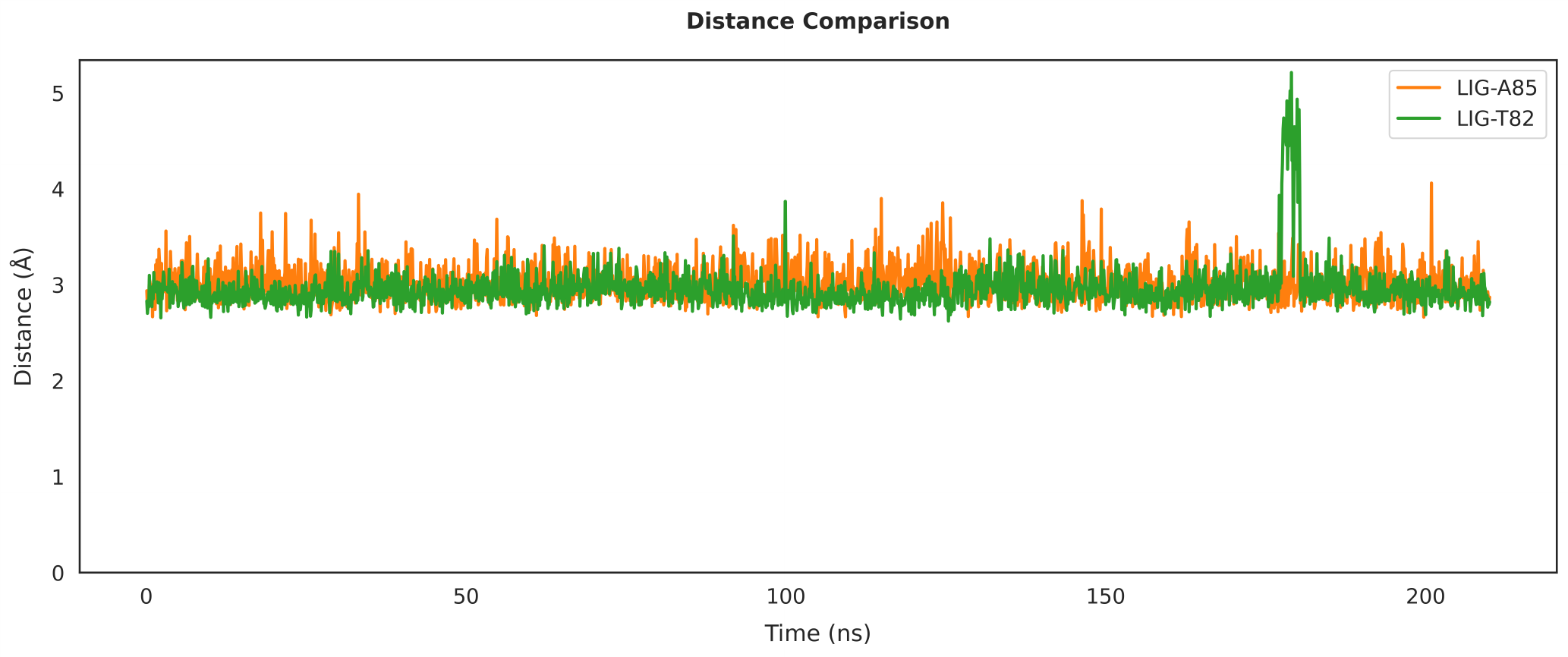}
    \caption{The distance plot between THR82 (green) and ALA85 (orange) and respective atoms of the ligand.}
    \label{fig:HBond}
\end{figure}
\subsubsection*{Seq2Seq1459}
The RMSD of the protein backbone, Seq2Seq1459 , and the ligand binding site exhibits significant fluctuations while remaining below 2.0 Å, notably, ligand RMSD shows the highest fluctuations throughout the simulation, with occasional spikes exceeding 2.4 Å, highlighting the inherent flexibility and mobility of the ligand. In contrast, the ligand binding site demonstrates lower RMSD values compared to the overall protein, indicating that the binding site is relatively more stable and less prone to large conformational changes. This observation is crucial, as it suggests that the binding site maintains its structural integrity despite the dynamic nature of the surrounding protein environment. Overall, the RMSD analysis reveals a balance between global protein flexibility and localized stability at the binding site, which is essential for maintaining functional interactions with the ligand. The RMSD plot is presented in Figure 8.
The hydrogen bond distance comparison plot (Figure 9) illustrates the temporal fluctuations in the distance between LIG-A85, key interaction for ligand stability. The LIG-A85 distance fluctuates within a range of approximately 3.0 - 3.5 Å throughout the simulation. The fluctuations suggest the dynamic behavior of this interaction, consistent with typical hydrogen bonding patterns observed in bio-molecular systems.
\begin{figure}
    \centering
    \includegraphics[width=1\linewidth]{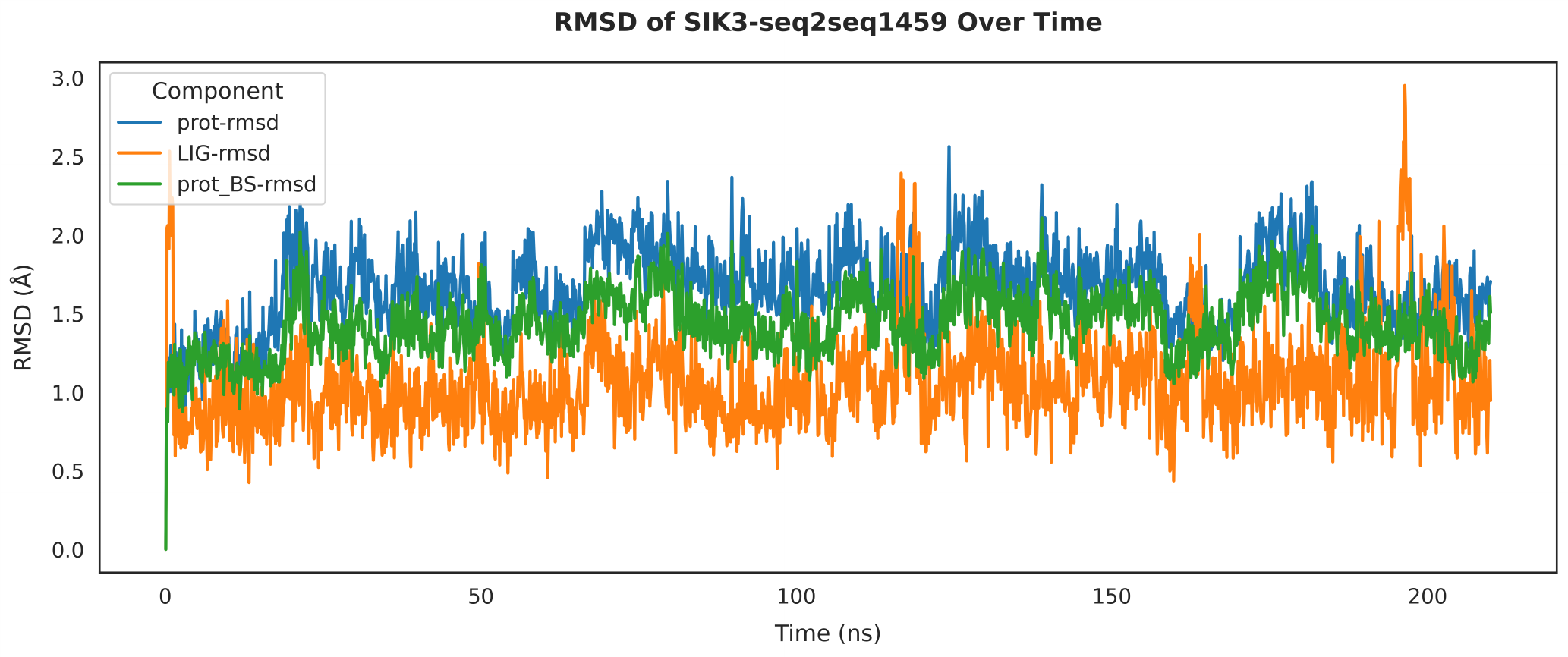}
    \caption{The RMSD plot of protein backbone (blue), Seq2Seq1459 ligand (orange), and ligand binding site (green).}
    \label{fig:RMSD}
\end{figure}
\begin{figure}
    \centering
    \includegraphics[width=1\linewidth]{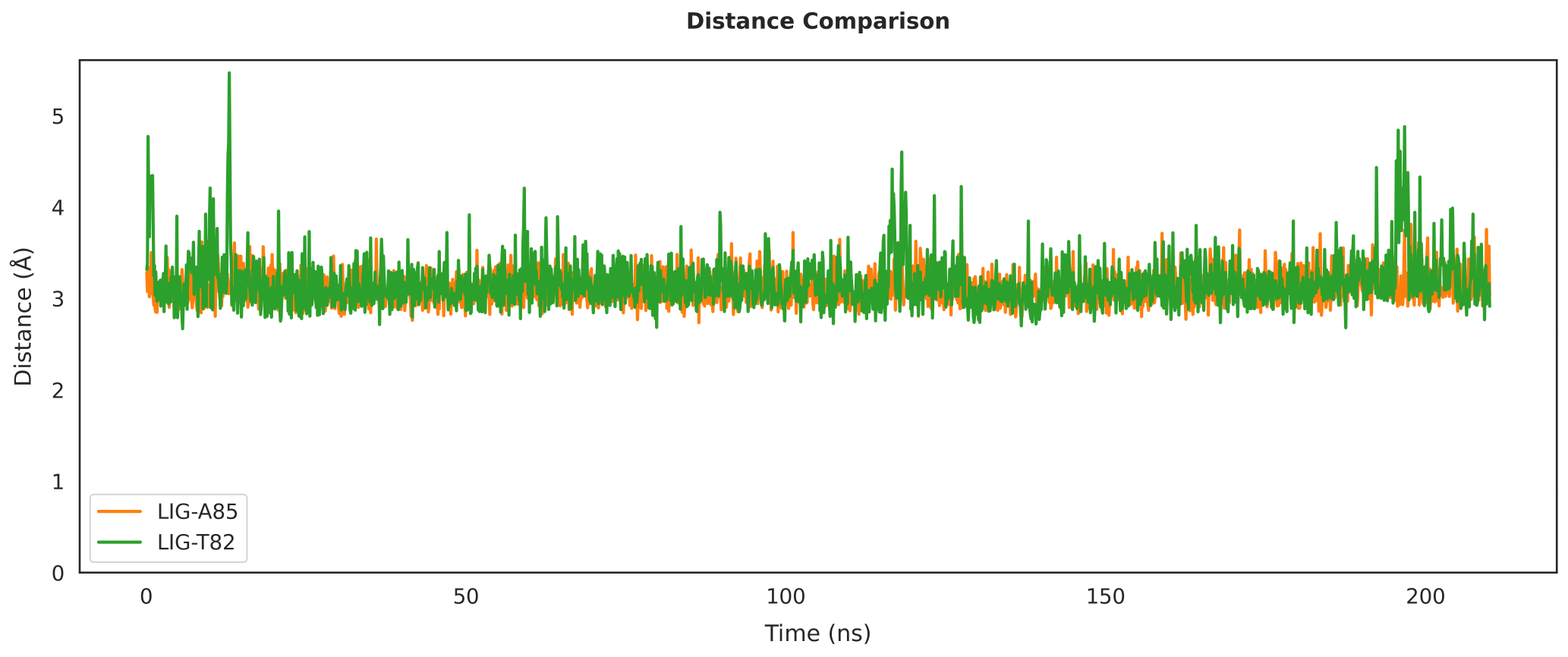}
    \caption{The distance plot between THR82 (green) and ALA85 (orange) and respective atoms of the ligand.}
    \label{fig:HBond}
\end{figure}

\subsubsection*{Seq2Seq2913}
The RMSD of the protein backbone shows stability around 1.2 - 1.8 Å, with periodic fluctuations reflecting moderate conformational changes. Similarly, the Seq2Seq2913 RMSD remains stable at a lower value (~1.2 - 1.8 Å), however small and short spikes are observed, suggesting that the ligand's flexibility. Notably, the RMSD of the ligand binding site residues remained consistently low throughout the simulation, stabilizing near 1.0 - 1.5 Å, indicating high rigidity and minimal deviation from the reference structure. This suggests that the binding site residues maintain a stable conformation, which is crucial for maintaining the protein-ligand interaction. The RMSD plot is presented in Figure 10.
The hydrogen bond distance comparison plot (Figure 11) illustrates the temporal fluctuations in the distances between key residues involved in hydrogen bonding interactions over the course of the simulation. The LIG-T82 interaction shows more pronounced peaks (3.1 - 4 Å), indicating disruptions and transient breaks in the hydrogen bond, whereas the LIG-A85 interaction appears more stable remaining close to 3.0 Å. These fluctuations suggest that LIG-A85 interaction maintains the overall stability, consistent with typical hydrogen bonding patterns observed in bio-molecular systems.
\begin{figure}
    \centering
    \includegraphics[width=1\linewidth]{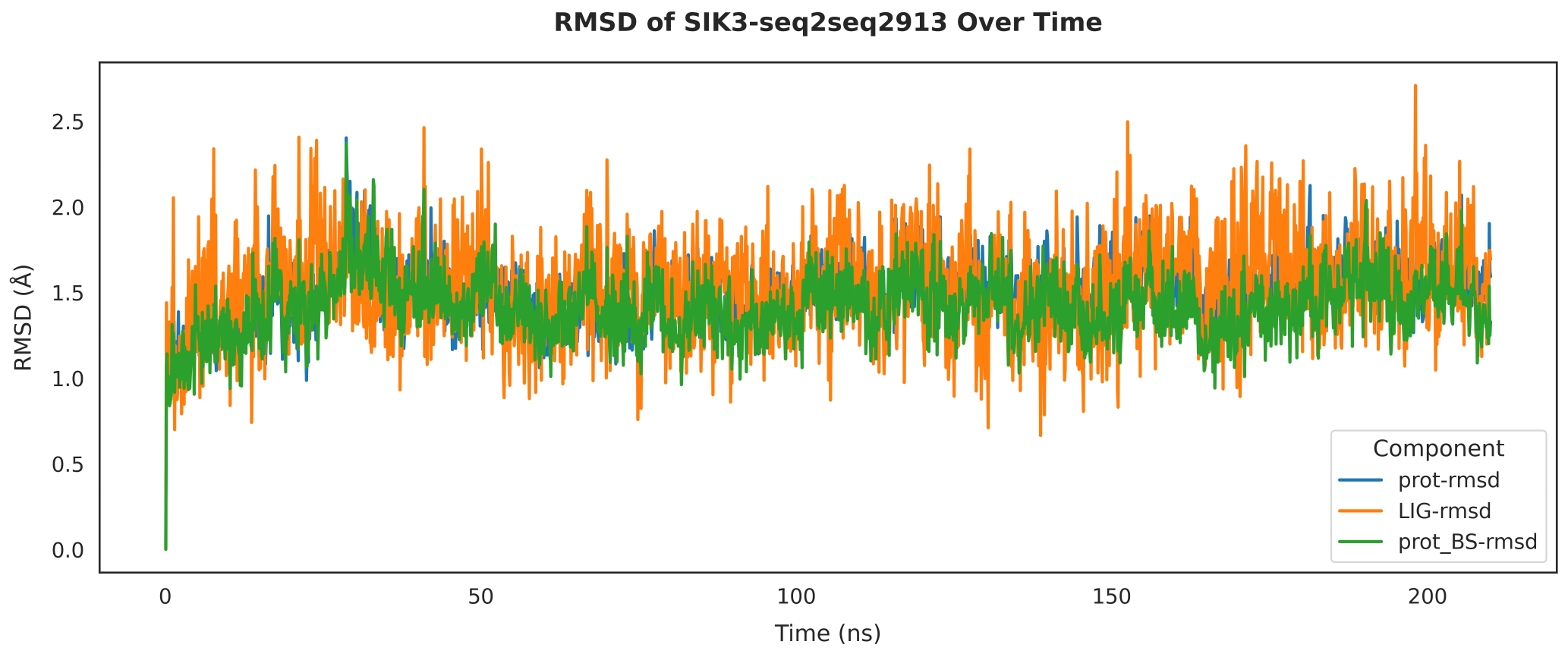}
    \caption{The RMSD plot of protein backbone (blue), Seq2Seq2913 ligand (orange), and ligand binding site (green).}
    \label{fig:RMSD}
\end{figure}
\begin{figure}
    \centering
    \includegraphics[width=1\linewidth]{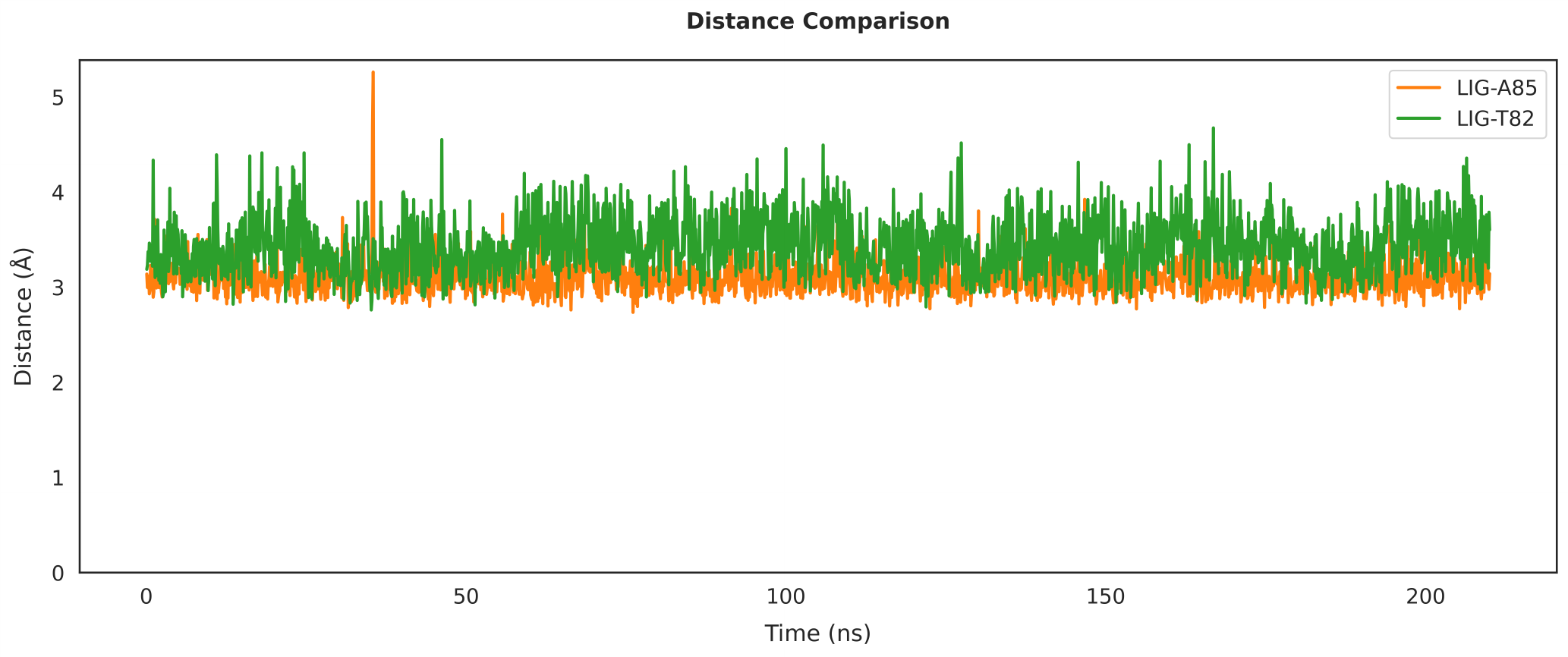}
    \caption{The distance plot between THR82 (green) and ALA85 (orange) and respective atoms of the ligand.}
    \label{fig:HBond}
\end{figure}
\subsubsection*{Seq2Seq3481}
The RMSD of the protein backbone and ligand binding site remained stable (~1.0 - 1.5 Å) throughout the simulation's time period with minimal fluctuations in the first 70 ns. In contrast, the Seq2Seq3481 RMSD shows significant fluctuations until 120 ns (~0.5 - 2.0 Å), suggesting that the ligands exhibited considerable movement during this time. However, the ligand achieved stability afterwards and remained stable till the end with RMSD ranging between ~2.0 - 2.2 Å. Notably, the RMSD of the ligand binding site residues remained consistently below the ligand and protein backbone RMSD throughout the simulation indicating high rigidity and minimal deviation from the reference structure. This suggests that the binding site residues maintain a stable conformation, which is crucial for maintaining the protein-ligand interaction. The RMSD plot is presented in Figure 12.
The hydrogen bond distance comparison plot (Figure 13) illustrates minimal fluctuations in the distances between key residues involved in hydrogen bonding interactions over the course of the simulation. Both interactions exhibit dynamic behavior, with distances fluctuating within a range of approximately 2.8 - 3.2 Å throughout the simulation. Both the LIG-A85 and LIG-T82 interaction appears stable, with negligible excursions. These fluctuations suggest that both interactions maintain overall stability, consistent with typical hydrogen bonding patterns observed in bio-molecular systems.
\begin{figure}
    \centering
    \includegraphics[width=1\linewidth]{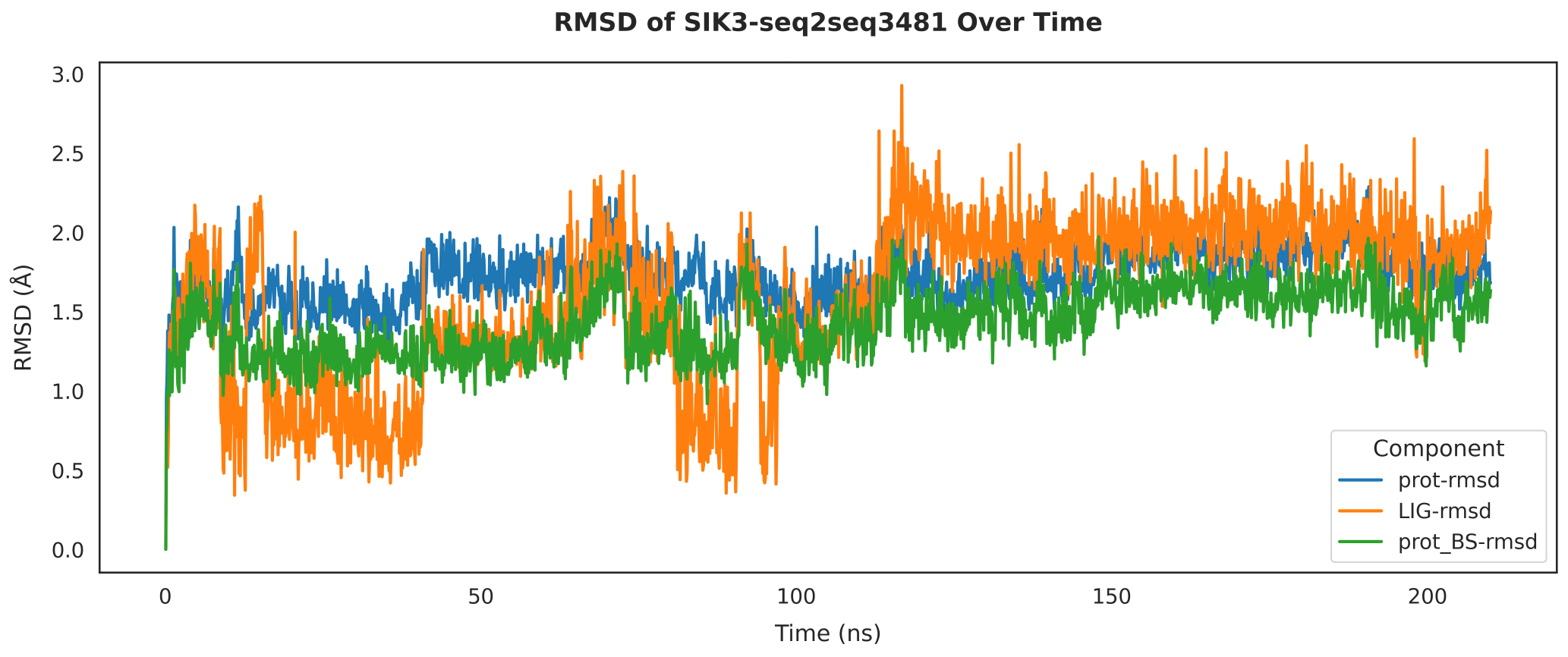}
    \caption{The RMSD plot of protein backbone (blue), Seq2Seq3481 ligand (orange), and ligand binding site (green).}
    \label{fig:RMSD}
\end{figure}
\begin{figure}
    \centering
    \includegraphics[width=1\linewidth]{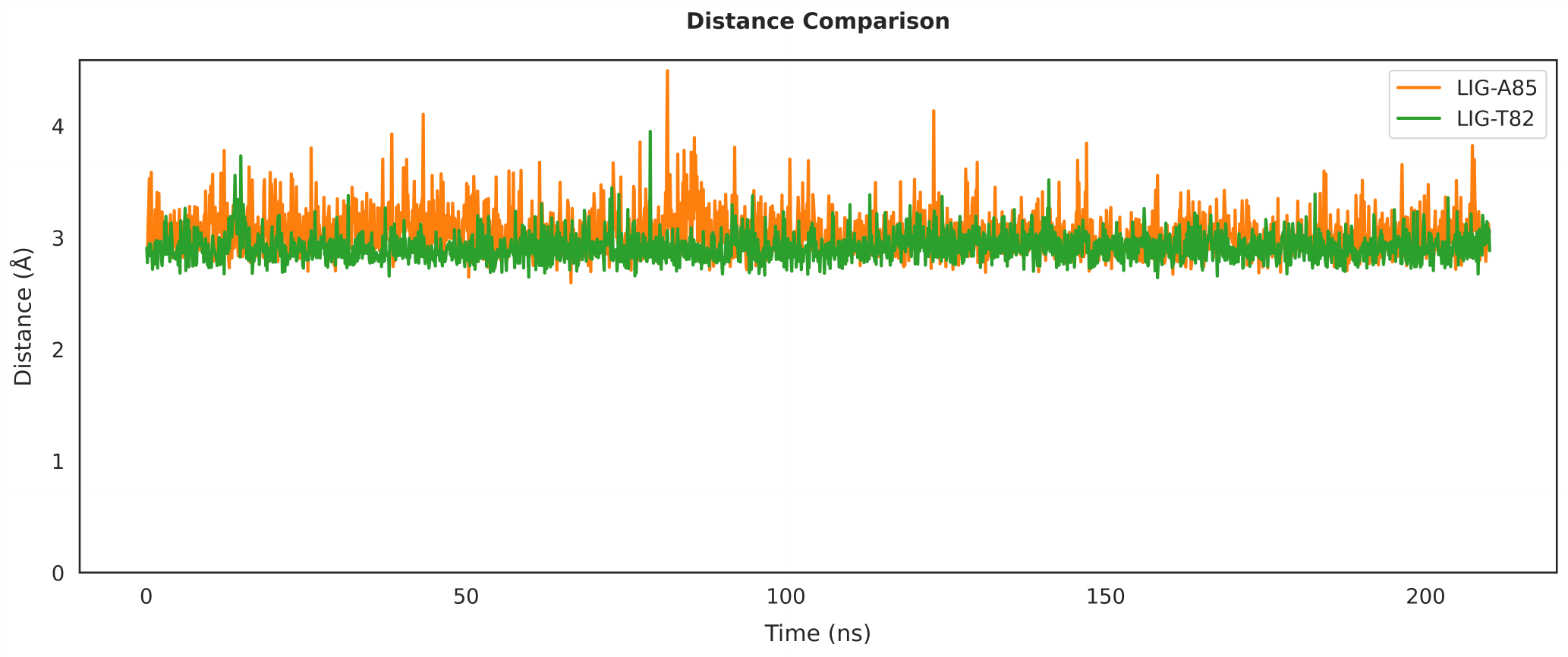}
    \caption{The distance plot between THR82 (green) and ALA85 (orange) and respective atoms of the ligand.}
    \label{fig:HBond}
\end{figure}
\subsubsection*{The Root Mean Square Fluctuations}
The RMSF (Root Mean Square Fluctuation) plot for the protein backbone (Figure 14) reveals distinct regions of structural flexibility across the sequence. Notably, several peaks are observed, indicating residues with high mobility. These peaks are particularly pronounced near residues 40, 220, and 260, suggesting that these regions experience significant conformational changes during the simulation. Additionally, the ligand binding site residues, marked by red dots, show relatively lower RMSF values compared to the surrounding regions, implying a more stable structure at the binding interface. This observation is consistent with the functional requirement for stability in ligand-binding pockets, which often necessitates reduced flexibility to maintain specific interactions. Overall, the RMSF analysis provides insights into the dynamic behavior of the protein, highlighting both flexible and rigid domains that may play critical roles in its function and interaction dynamics.
\begin{figure}
    \centering
    \includegraphics[width=1\linewidth]{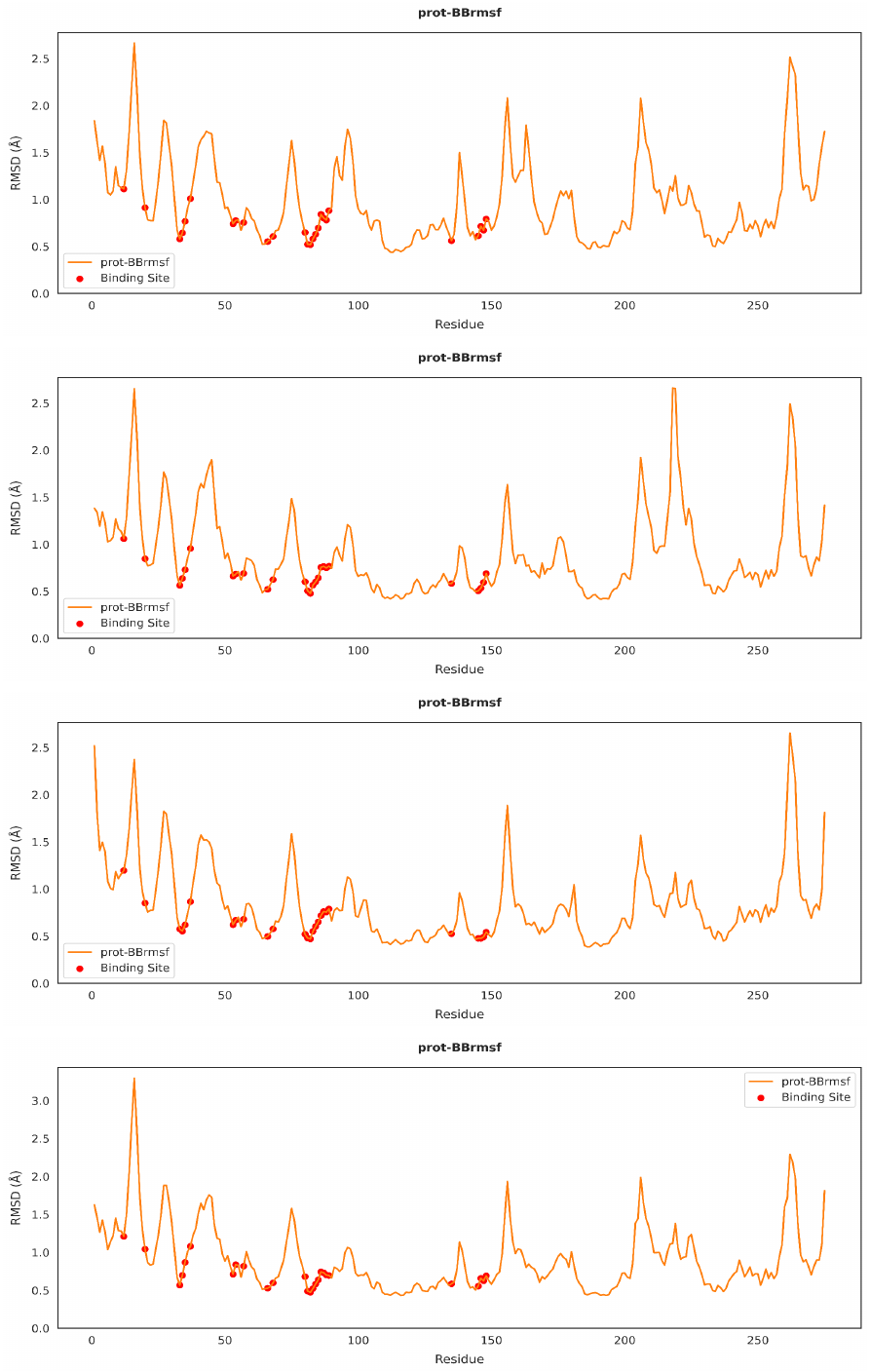}
    \caption{The RMSF plot of protein backbone of protein-ligand complexes in order of, SIK3-1030, SIK3-1459, SIK3-2913 and SIK3-3481, respectively}
    \label{fig:enter-label}
\end{figure}
\subsection{AiZynthFinder Retro-synthesis}
The all atom molecular dynamics simulations validated the stability of the protein-ligand complexes, AiZynthFinder, a retro-synthesis model was used to predict the synthesis path and the purchasable precursors available in ZINC database for molecules 1030, 1459, 2913 and 3481. The easiest to synthesize is 1030 with state scores of 0.9976 and a two step synthesis reaction with purchasable substrates available in ZINC, Figure 15. The reaction synthesis pathways for each molecules and scores are provided in supplementary Figure 1 and supplementary Table 1. 
\begin{figure}
    \centering
    \includegraphics[width=1\linewidth]{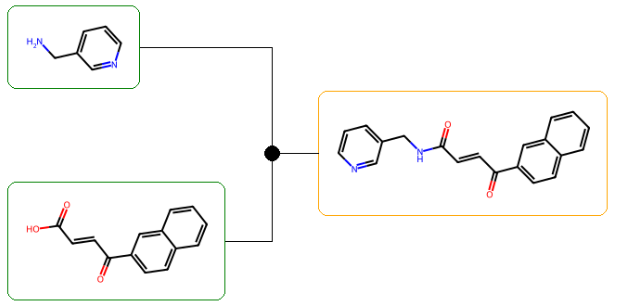}
    \caption{The purchasable precursors identified for Seq2Seq-1030 molecules by AiZynthFinder}
    \label{fig:placeholder}
\end{figure}

\subsection{Random Selection ChEMBL}
A set of 100 molecules were randomly selected from ChEMBL database for docking with SIK3, allowing us to estimate the molecules generated by Seq2Seq-VAE is better than just random chance selection from ChEMBL. The Ala145 hydrogen bond constraint and filtering for existence of Thr142 hydrogen bond was used as a criteria to asses suitability of the molecules. Only one molecules Chembl200118 passed both docking criteria with a docking score $-7.5 \, \text{kcal/mol}$, Figure 16. 
\begin{figure}
\centering
    \includegraphics[width=0.5\linewidth]{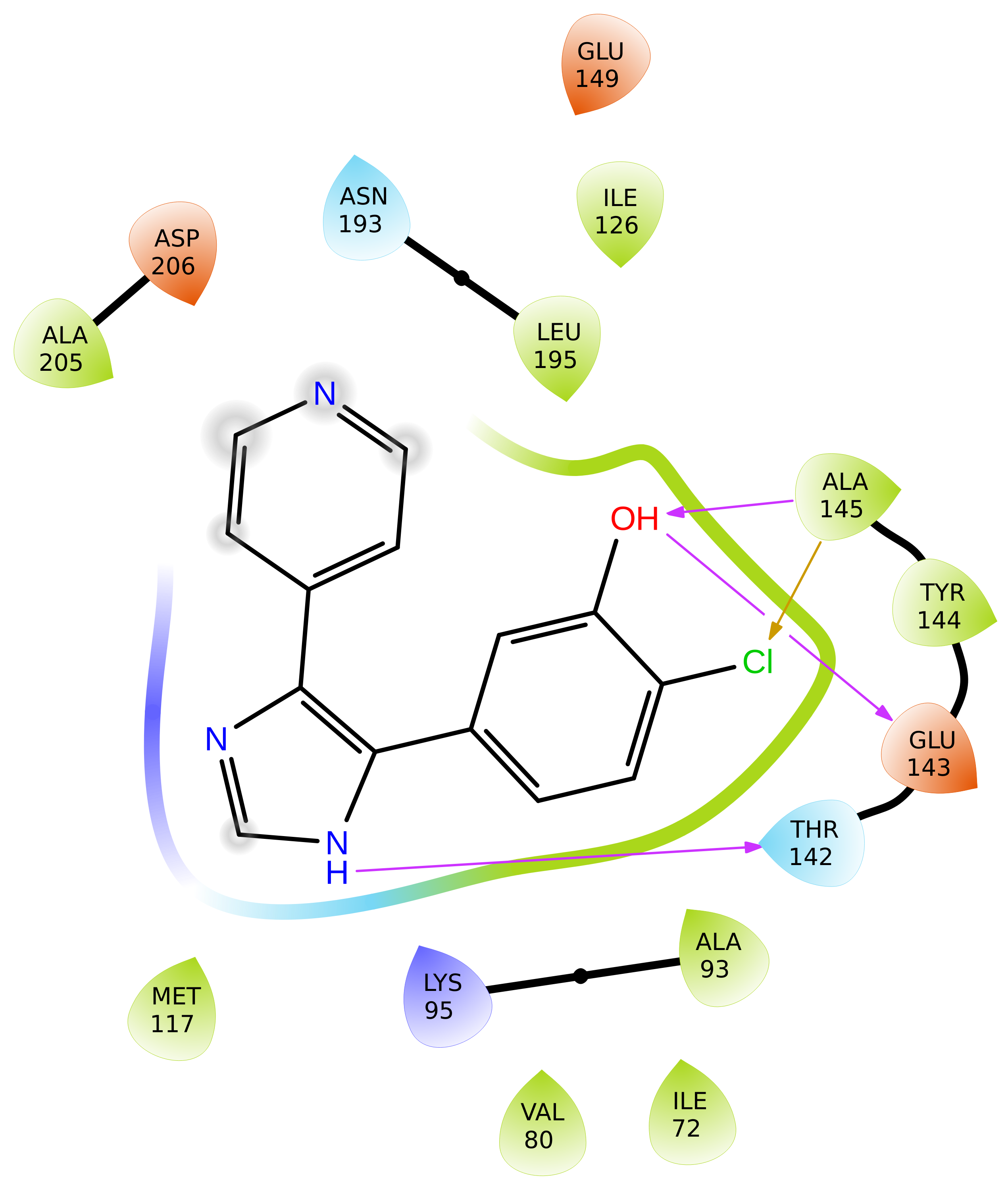}
\caption{Chembl200118, docking score  $-7.5 \, \text{kcal/mol}$ interaction with SIK3 ligand binding site exhibiting Ala145 and Thr142 hydrogen bonds}
    \label{fig:placeholder}
\end{figure}

\subsection{DiffSBDD}
DiffSBDD, a structure-based drug design (SBDD) model leveraging SE(3)-\linebreak equivariant 3D conditional diffusion to generate drug-like molecules conditioned explicitly on protein binding pockets. It takes into account the spatial orientation of the ligand binding pocket, it was used to generate 100 molecules providing SIK3-1030 as reference protein-ligand complex. The DiffSBDD generated 82 valid molecules where 12 molecules passed the docking score cutoff of $\leq -7.0 \, \text{kcal/mol}$. Table 2 provides the top 4 candidate molecules generated by DiffSBDD with docking scores.

\begin{table}
\centering
\caption{DiffSBDD generated molecules with docking score lower than  $-7.0 \, \text{kcal/mol}$ exhibiting the Ala145 and Thr142 hydrogen bonds critical for ligand stability and SIK3 selectivity, respectively}
\label{tab:placeholder}
    \begin{tabular}{|c|c|}\hline
         DiffSBDD mol& Docking score\\\hline
         mol69& -7.86\\\hline
         mol94& -7.71\\\hline
         mol33& -7.64\\\hline
         mol98& -7.62\\\hline
         mol67& -7.15\\ \hline
    \end{tabular}

\end{table}

\subsection{Off Target Selectivity}
Off target binding is a critical hurdle in drug discovery where a molecules binds to unintended target proteins leading to undesirable effects. AMP-activated protein kinase (AMPK) (pdb id; 4rer),  Microtubule-associated protein/Microtubule affinity-regulating kinase 4 (MARK4) (pdb id; 5es1) and NUAK family SnF1-like kinase-1 (NUAK1) (pdb id; 8oui) were identified to have similar ligand binding site profile to SIK3. NUAK1 and MARK4 has Ala135 and Met132 corresponding to Ala145 and Thr142 in SIK3, the Ala135 was used as a hydrogen bond constraints in docking protocol as was Ala145 in SIK3 docking. The docking score of the selected molecules with all three kinases shows the specificity of molecules for SIK3 kinases as none of the molecules exhibit better docking score for any of off target kinases than SIK3. this reaffirms the SIK3 specificity of the molecules generated by Seq2Seq-VAE model. Table 3 provides the docking score of selected molecules for each kinase.

\begin{table}
    \centering
\caption{The docking score of selected molecules with MARK4, NUAK1 and AMPK with respective docking scores}
\label{tab:my_table}
    \begin{tabular}{|c|c|c|}\hline
         Kinase&  Mol ID& Docking Score\\\hline
         MARK4 (5es1)&  Seq2Seq3481& -7.46\\\hline
         &  Seq2Seq2913& -5.58\\\hline
         &  Seq2Seq1459& -5.25\\\hline
         &  Seq2Seq1030& -3.90\\\hline
         NUAK1 (8oui)&  Seq2Seq2913& -6.85\\\hline
         &  Seq2Seq1459& -6.44\\\hline
         &  Seq2Seq3481& -5.14\\\hline
         &  Seq2Seq1030& -4.47\\\hline
         AMPK (4rer)&  Seq2Seq1459& -7.19\\\hline
 & Seq2Seq1030&-7.03\\\hline
 & Seq2Seq2913&-6.84\\\hline
 & Seq2Seq3481&-6.01\\ \hline
    \end{tabular}

\end{table}

\section*{Discussion}
The development of effective therapeutics for AD remains a challenge due to its complex pathophysiology and the limitations of traditional drug discovery approaches \cite{Fang2020-ca}. We developed a \textit{de novo} drug design pipeline integrating a Seq2Seq-VAE model with a two-step AL workflow to generate molecules with optimized physio-chemical properties and high binding affinity for SIK3, an emerging therapeutic target implicated in AD-related circadian rhythm disruptions \cite{Dai2024-ew}.  While taking a similar approach to \cite{filella2025optimizing}, our results demonstrate the efficacy, scalability, and robustness of this method, as evidenced by the progressive enrichment of the chemical space with SIK3-specific, BBB permeable molecules, offering a promising framework for addressing underexplored targets where available data is scarce. The Seq2Seq-VAE model for \textit{de novo} drug design combines a sequence-to-sequence architecture with a variational autoencoder to learn a latent representation of molecular structures. The encoder maps SMILES into a continuous latent space and the decoder reconstructs valid molecular sequences from sampled latent vectors. The VAE in this model introduce a probabilistic component to the latent space, allowing for the generation of diverse and realistic molecular structures through the reparameterization trick. Additionally, the VAE encourages the latent space to be smooth and continuous, facilitating the exploration of chemical space and the generation of novel molecules with desired properties. The model is trained using a combination of reconstruction loss and Kullback-Leibler (KL) divergence to balance structural accuracy and latent space regularization \cite{Yoshikai2024-mv, Kingma2019-cl}. \par

The AL workflow proved highly effective in generating molecules with desired pharmacological profiles where the inner loops, focused on optimizing physio-chemical properties, leveraged stringent filters based on established descriptors, such as QED, SA, and BBB permeability criteria (Table 1). The progressive tightening of QED cutoffs (from $\geq 0.5$ to $\geq 0.6$) and Tanimoto similarity (from 0.8 to 0.7) across the inner loops was a deliberate strategy to balance exploration and exploitation, allowing the model to refine its output while maintaining structural diversity. This is reflected in the increase in molecules meeting physio-chemical properties criteria, from 56\% in early cycles to 85.4\% in later cycles, despite a slight reduction in valid molecule yields (from 91\% to 71.6\%). The use of a KDE-based sampling strategy with a bandwidth of 0.3 enhanced efficiency by focusing the model on high-density regions of the latent space associated with desirable properties, justifying its selection over broader sampling methods that might dilute target-specific optimization. \par

The outer loops, designed to enhance SIK3 binding affinity, utilized Glide docking scores with cutoffs of $\leq -7.0 \, \text{kcal/mol}$ (cycles 1--4) and $\leq -7.5 \, \text{kcal/mol}$ (cycles 5--8). These thresholds were chosen based on standard practices in molecular docking, where scores below $-7.0 \, \text{kcal/mol}$ indicate strong binding affinity, and the stricter $-7.5 \, \text{kcal/mol}$ cutoff in later cycles aimed to prioritize high-potency candidates. The growth of the permanent-specific set from 103 to 3,445 molecules, with the proportion of molecules meeting these criteria increasing from first outer loop cycle to the last outer loop cycle, underscores the workflow’s ability to iteratively steer the model toward a functionally relevant chemical space. The Ala145 hydrogen bond constraint docking accurately replicate realistic environment, as this residue is critical for stable ligand interactions in the kinase hinge region. Furthermore, the additional filtering for Thr142 hydrogen bonds, informed by Galapagos’ findings on SIK selectivity \cite{Oster2024-aa, Temal-Laib2024-gf}, enhanced the specificity of generated molecules for SIK kinases over related families like AMPK, addressing a key challenge in kinase inhibitor design. \par

The molecular docking provides critical insights into the ligand-protein interactions however it lacks the details of a cellular environment and might often mislead. To assist the results of molecular docking we performed Molecular dynamics (MD) simulations for selected molecules filtered from the docking results. MD simulations revealed distinct dynamic behaviors across the designed ligands, with key insights into binding stability and residue-specific interactions. All systems achieved equilibrium after initial relaxation, with backbone RMSD values stabilizing below 2.0 Å, indicative of overall structural integrity. Notably, ligand binding sites exhibited consistently lower RMSD and RMSF values compared to global protein fluctuations, underscoring their rigidity, a critical feature for maintaining productive ligand interactions. Hydrogen bond analysis highlighted the stability of the LIG-A85 interaction (critical for maintaining ligand stability), while LIG-T82 showed greater variability, suggesting its role in modulating SIK family specificity. Seq2Seq1030 and Seq2Seq3481 demonstrated particularly stable binding profiles, with ligand RMSD converging to sub-2.0 Å ranges after equilibration. In contrast, Seq2Seq1459 exhibited higher ligand mobility, potentially reflecting suboptimal packing. These findings collectively validate the stability of ligands within the binding site of SIK3 while providing important insights into ligand dynamics allowing space for ligand optimization that may influence affinity and selectivity. Further ligand optimization could target flexible regions (e.g., residues 40, 220, 260) to enhance conformational stability.

A critical aspect of our workflow is its ability to maintain scaffold diversity, preventing mode collapse, a common pitfall in generative models. The observed decline in scaffold diversity (from 0.81 to 0.65 in inner loops and 0.91 to 0.61 in outer loops) reflects a controlled convergence toward optimized scaffolds, yet diversity remained robust, ensuring exploration of varied chemical space regions. This balance was achieved through the Tanimoto similarity cutoff (0.7-0.8), which discarded structurally redundant molecules, and the iterative updating of the temporal-specific set, which enriched the training data with diverse, high-quality candidates. The UMAP visualization (Figure 4) further confirms efficient chemical space exploration, validating the model’s learning capacity. \par

The reliability of our results is supported by the significant increase in molecules passing both physio-chemical and docking filters, coupled with the model’s performance despite limited initial SIK3-specific data (148 molecules from PubChem) \cite{Kim2016-xe}. This scarcity, typical for novel targets, highlights the scalability of our AL framework, which iteratively refines the latent space without requiring large labeled datasets. The Seq2Seq-VAE’s ability to learn from a general ChEMBL dataset ($\sim 670$k SMILES) and adapt to SIK3-specific requirements through fine-tuning demonstrates its versatility, making it applicable to other underexplored targets in AD and beyond. \par

The workflow’s two-step AL approach offers a generalizable strategy that can be adapted to other protein targets with minimal modifications, addressing the bottleneck of molecular design in early-stage drug discovery. While the generated molecules require experimental validation, their high docking scores and optimized physio-chemical properties profiles suggest strong potential as SIK3 inhibitors, particularly for restoring circadian rhythm disruptions in AD. This study establishes a robust and scalable pipeline for \textit{de novo} drug design, leveraging the synergy of Seq2Seq-VAE and AL to generate SIK3-targeted molecules with therapeutic promise. The rational selection of parameters and thresholds, grounded in established chemical and pharmacological principles, ensures the reliability and reproducibility of our results. By demonstrating the feasibility, our work paves the way for further exploration of circadian rhythm-based therapies and underscores the transformative potential of generative AI in accelerating drug discovery for complex diseases. \par

\section*{Methods}
\subsection*{General and SIK3 Specific Training set}
The general training set was sourced from the ChEMBL compound database \cite{Gaulton2012-pk}. To generate a focused and relevant dataset, SMILES representations underwent a systematic filtering. First, SMILES strings were standardized to ensure uniformity, and duplicates were removed to eliminate bias or over-representation of specific compounds in the training data. Subsequent refinement involved applying character-length filters: SMILES shorter than 15 characters or longer than 80 characters were excluded, as these likely correspond to overly simplistic or excessively complex molecules, potentially hindering BBB permeability or synthetic feasibility. The resulting general training set comprised ($\sim 670$k unique SMILES), encompassing a diverse array of compounds with drug-like properties. \par

For target-specific fine-tuning of the pre-trained model, a dataset focused on SIK3 was constructed. Molecules reported as active against SIK3 were retrieved from the PubChem database \cite{Kim2025-me, Kim2016-xe}, a public repository of chemical substances and their biological activities. These molecules were subjected to a filtering process based on character length, retaining only SMILES ranging from 15 to 80 characters, maintaining consistency with the general training set’s design. The final SIK3-specific dataset consisted of 148 molecules meeting this filtering criterion, providing a targeted set for refining the model’s predictive capabilities toward SIK3-related pharmacological profiles. \par

\subsection*{Training of Seq2Seq-VAE model}
A Sequence to Sequence Variational Autoencoder (Seq2Seq-VAE) model was trained on the general dataset to learn the underlying pattern of molecules and return a low dimensional latent space that can be decoded to generate novel drug-like molecules. The Seq2Seq-VAE model for \textit{de novo} drug design combines a sequence-to-sequence (Seq2Seq) architecture with a variational autoencoder (VAE) to learn a latent representation of molecular structures. The encoder, a single layer, 256 units, bidirectional Long Short Term Memory (LSTM), maps SMILES into a continuous latent space, while a single layer, 256 units, LSTM decoder reconstructs valid molecular sequences from sampled latent vectors. The VAE component introduces a probabilistic latent space, enabling generative molecular design via the reparameterization trick. The model optimizes a loss function combining reconstruction error and KL divergence regularization (enforcing latent space continuity), striking a balance between precise molecule generation and exploratory capacity \cite{Yoshikai2024-mv, Kingma2019-cl}. \par

Tokenization of the focused general training set and SIK3 specific set was performed at the character level, incorporating special tokens (G and E for start and end/padding, respectively) to denote sequence boundaries. To standardize input dimensions, sequences were padded to a predefined length of 80 characters maximum, enabling efficient learning within the model. The dataset was transformed into a one-hot encoded matrix where the input sequences were shifted left by one character to generate target sequences for supervised learning. During inference, latent vectors were sampled from the learned distribution and decoded into novel molecular structures, enabling \textit{De novo} molecular generation with controlled diversity and validity. \par

\subsection*{Active Learning (AL) Workflow}
In the target-specific fine-tuning phase, an AL workflow was employed to iteratively guide the Seq2Seq-VAE model towards generating molecules with desired properties. In this workflow a two stage approach of guiding the model towards generating target specific molecules with desired properties was adopted. The inner loop generates a set of SMILES after training for a given number of epochs which are subsequently filtered for physio-chemical properties (validity, QED, and SA). Validity checks ensure the molecules are chemically viable by converting the SMILES string to molecules using the RDKit library la\cite{Landrum2016RDKit2016_09_4} whereas SA estimates the difficulty level of molecule synthesis. The SA score ranges from 0 to 10 where a lower score indicates ease of synthesis and vice versa, SA score cutoff $\leq 6$ was used to prioritize molecules. The process was repeated iteratively, allowing the model to converge toward generating molecules with enhanced desired properties while maintaining structural diversity and chemical validity. \par 

A directed strategy was implemented to sample the chemical space focusing on molecules that satisfy predefined physio-chemical property filters. Kernel Density Estimation, a non-parametric statistical method, was utilized to explore this chemical space and steer the optimization process toward regions enriched with molecules exhibiting desired physio-chemical profiles. Latent representations were derived from molecules that satisfied the physio-chemical property filters. KDE was subsequently employed to estimate the probability density function (PDF) of these latent representations, facilitating the identification of high-density regions corresponding to desirable chemical properties. The resulting PDF was used for sampling 1500 new latent vectors, prioritizing the generation of molecules with optimized physio-chemical profiles. The generated molecules were subsequently filtered using the physio-chemical constraints in Table 1. By incorporating KDE, we aimed to effectively bias the molecular generation toward structurally and functionally relevant compounds, enhancing the specificity and efficiency of the molecular design process. \par

\subsection*{Docking Protocol}
The molecular docking of molecules from the temporal-specific set, a set of molecules collected during each inner loop cycle, with SIK3 was carried out using the GLIDE tool from the Schrödinger software suite (Schrödinger Release 2024-1, Schrödinger, LLC, New York, USA) with the standard precision protocol (SP) \cite{Yang2021} \cite{Halgren2004} \cite{Friesner2004}. The molecules were prepared with LigPrep, setting the pH value to $7.4 \pm 0.5$ and generating a maximum of four tautomers for each molecule. The crystal structure of SIK3 was obtained from the RCSB Protein Data Bank (PDB code: 8r4v), and hydrogen atoms were added using the Protein Preparation Wizard tool. A docking grid box of 29 Å with X, Y, and Z coordinates $-1.90$, $-60.06$, and $-4.14$, respectively, was generated using the SIK3 co-crystallized ligand as the center of the box, and a hydrogen bond constraint with the backbone NH of the hinge residue (Ala145) was applied. The Ala145 hydrogen bond is critical for stability of ligand in the hinge region of SIK3. \cite{T\cite{}emal-Laib2024-gf} \par

\subsection*{Molecular Dynamics Simulations of Selected Candidates}
All-atom molecular dynamics (MD) simulations of the selected protein-ligand complexes were performed using the Amber software suite \cite{Case_AmberTools}. Ligand parameters were derived from quantum mechanical calculations using Jaguar \cite{bochevarov2013jaguar}, including geometry optimization at the B3LYP/6-31G(d) level and PCM solvation model-based partial charges; the General Amber Force Field (GAFF) \cite{wang2004development} was applied for ligand parametrization. Protein systems were prepared with PDB4amber to optimize protonation states and remove redundant atoms. Topology and coordinate files were generated in tLeap using the ff14SB \cite{maier2015ff14sb} force field for proteins, TIP3P water molecules, and Na+/Cl- ions to neutralize system charge. Each system was subjected to two-stage energy minimization: first restraining protein and ligand atoms (residues 1–276, 50 kcal/mol/Å²), followed by release of hydrogen atoms in the second stage with restraints on non-hydrogen atoms of the protein-ligand complex only. Systems were then heated over three stages from 100K to 300K under constant volume, applying positional restraints, and transitioning to constant pressure (1 atm) in the final stage. Subsequent equilibration involved stepwise release of protein restraints (residues 1–275, 50 to 0.5 kcal/mol/Å²) while maintaining ligand restraint (residue 276, 50 kcal/mol/Å²), followed by progressive release of the ligand (50 to 0.1 kcal/mol/Å²) with weak terminal residue restraints (residues 1 and 275, 0.5 kcal/mol/Å²). Finally, production simulations were carried out under NPT conditions (300 K, 1 atm) with isotropic pressure coupling, a 2 fs time step, and SHAKE \cite{krautler2001fast} constraints for hydrogen bonds. A total of 200 ns simulations were performed per system, with coordinates saved every 100 ps for analysis.

\subsection{Comparative analysis}
AiZynthFinder, a retrosynthetic model that employs a Monte Carlo tree search (MCTS) algorithm guided by a neural network policy trained on known reaction templates \cite{genheden2020aizynthfinder}. AiZynthFinder recursively breaks down molecules into purchasable precursors by proposing chemical disconnections step-by-step, enabling synthesis route construction for complex molecules. In our work, AiZynthFinder was utilized to predict synthetic routes for selected \textit{de novo} molecules generated by Seq2Seq-VAE model, thereby reinforcing that molecule selection aligns with synthetic accessibility scores. This step reaffirms that the molecules generated and selected for SIK3 targeting are not only theoretically effective but also practically synthesizable, which is crucial for downstream drug development. \par 

A set of 100 molecules, randomly selected, from the ChEMBL database were docked to SIK3 with exact parameters. The Seq2Seq-VAE model’s output had been vetted through a combination of \textit{in silico} validation filters, which ensures that generated molecules meet certain drug-like and binding-relevant properties. This benchmarking step provides a validation baseline by comparing how well randomly selected molecules from a comprehensive public database perform relative to the molecules from our generative model finetuned through the active learning pipeline. \par 

Further, a generative model, DiffSBDD, a structure-based drug design (SBDD) model leveraging SE(3) - equivariant 3D conditional diffusion to generate drug-like molecules conditioned explicitly on protein binding pockets was used to generated 100 molecules \cite{schneuing2024structure}. DiffSBDD respects important spatial symmetries, ensuring physically plausible molecular structures in the 3D binding environment. For evaluation, DiffSBDD was provided with the SIK3 and molecules generated by the Seq2Seq-VAE model, generating new candidate molecules. This setup further verify that the Seq2Seq-VAE generated molecules not only score well in traditional docking but also outperform an SBDD methods that incorporate spatial binding pocket information and geometric symmetries. \par 

Off-target binding, a major hurdle in targeted drug development, was investigated via embedding pairing from binding site representations extracted from PickPocket \cite{tarasi2025evolutionary}. All proteins from the kinase family were extracted from KLIFS \cite{kanev2021klifs}. A single PDB structure was used per kinase entry when a crystallographic structure was available. Else, a predicted structure from AlphaFold \cite{jumper2021highly}database was chosen only if the mean pLDDT was $\geq$ 60. Pickpocket was used to identify binding sites in all structures and to extract embeddings for all identified binding sites. Pairwise euclidean distances were computed for all embeddings. Those pairs with an embedding distance lower that 10 were selected as to have a similar binding site to SIK3. The selected molecules were docked against a panel of kinases to assess specificity and potential off-target interactions. This step is vital to understanding the selectivity profile of Seq2Seq-VAE generated molecules, ensuring that while they bind effectively to SIK3, they do not undesirably interact with other kinases that could cause side effects. Together, these four methods create a robust, multi-faceted framework for validating the performance and utility of the Seq2Seq-VAE model in generating target-specific, synthetically accessible, and biologically relevant molecules for SIK3.

\bibliographystyle{unsrt} % Sort citations by order of appearance
\bibliography{references}

\section*{Acknowledgments}
This rsearch work was supported by the “Targeting Circadian Clock Dysfunction in Alzheimer’s Disease” Doctoral Network (TClock4AD), a joint doctoral program funded within Horizon Europe Marie Skłodowska-Curie Doctoral Networks (grant agreement No. 101072895).

\section*{Author contributions statement}
Shah Zeb Khan is a Ph. D. student at the Nostrum Biodiscovery, Barcelona, Spain, is supported by the scholarship of the “Targeting Circadian Clock Dysfunction in Alzheimer’s Disease” Doctoral Network (TClock4AD), a joint doctoral program funded within Horizon Europe Marie Skłodowska-Curie Doctoral Networks.

S.K, A.M, C.P and J.I conceived the experiment(s),  S.K. conducted the experiment(s), S.K. and A.M. analysed the results.  All authors reviewed the manuscript. 

\begin{table}[ht]
\centering
\captionsetup{justification=justified, singlelinecheck=false}
\caption{Performance metrics of the Seq2Seq-VAE model across training epochs, evaluated during the active learning (AL) workflow for SIK3-specific molecule generation. The model generated 7500 SMILES per 25 epochs from KDE-sampled latent vectors (bandwidth 0.3) for epochs 25--100 and 15000 SMILES per 25 epochs for epochs 125--200. Property-filtered percentages reflect the proportion of KDE-generated SMILES meeting physio-chemical criteria (Table 1), while Glide-filtered percentages indicate the proportion of property-filtered molecules with docking scores $\leq -7.0 \, \text{kcal/mol}$ (epochs 25--100) and $\leq -7.5 \, \text{kcal/mol}$ (epochs 125--200).}
\label{tab:performance_metrics}
\begin{tabularx}{\textwidth}{|p{1.0cm}|p{0.8cm}|p{0.8cm}|p{1.2cm}|p{1.2cm}|p{2.0cm}|p{2.5cm}|}
\hline
\textbf{Epoch} & \textbf{Train Set} & \textbf{Test Set} & \textbf{KDE-gen SMILES} & \textbf{Prop Filtered} & \multicolumn{2}{c|}{\textbf{Internal Scaffold Diversity}} \\
\cline{6-7}
 &  &  &  &  & \textbf{Property Filtered} & \textbf{Glide Filtered} \\
\hline
25 & 103 & 35 & 6837 & 3849 (56.2\%) & 0.88 -- 0.81 & 0.87 -- 0.91 \\
50 & 314 & 35 & 6636 & 4490 (67.6\%) & 0.87 -- 0.78 & 0.87 -- 0.87 \\
75 & 648 & 35 & 6460 & 4938 (76.4\%) & 0.88 -- 0.77 & 0.87 -- 0.83 \\
100 & 1218 & 35 & 6209 & 5015 (80.7\%) & 0.87 -- 0.73 & 0.86 -- 0.73 \\
125 & 1512 & 35 & 11448 & 9054 (79\%) & 0.87 -- 0.69 & 0.86 -- 0.70 \\
150 & 2089 & 35 & 11309 & 9278 (82\%) & 0.87 -- 0.66 & 0.86 -- 0.67 \\
175 & 2734 & 35 & 11028 & 9283 (84.1\%) & 0.87 -- 0.66 & 0.86 -- 0.65 \\
200 & 3445 & 35 & 10746 & 9187 (85.4\%) & 0.87 -- 0.65 & 0.86 -- 0.61 \\
\hline
\end{tabularx}
\end{table}

\end{document}